\documentclass[aps,twocolumn,superscriptaddress]{revtex4}
\usepackage{graphicx}

\newcommand{\ket}[1] {\left| #1 \right\rangle}

\begin{document}

\title{Superradiance phase transition in the presence of parameter fluctuations}

\author{S. Ashhab}
\affiliation{Qatar Environment and Energy Research Institute, Hamad Bin Khalifa University, Qatar Foundation, Doha, Qatar}
\author{K. Semba}
\affiliation{National Institute of Information and Communications Technology, 4-2-1, Nukuikitamachi, Koganei, Tokyo 184-8795, Japan}

\begin{abstract}
We theoretically analyze the effect of parameter fluctuations on the superradiance phase transition in a setup where a large number of superconducting qubits are coupled to a single cavity. We include parameter fluctuations that are typical of superconducting architectures, such as fluctuations in qubit gaps, bias points and qubit-cavity coupling strengths. We find that the phase transition should occur in this case, although it manifests itself somewhat differently from the case with no fluctuations. We also find that fluctuations in the qubit gaps and qubit-cavity coupling strengths do not necessarily make it more difficult to reach the transition point. Fluctuations in the bias points, however, increase the coupling strength required to reach the quantum phase transition point and enter the superradiant phase. Similarly, these fluctuations lower the critical temperature for the thermal phase transition.
\end{abstract}

\maketitle

\section{Introduction}

Cavity quantum electrodynamics (QED), the study of the interaction between matter and the electromagnetic field inside a cavity at the quantum level, allows the investigation of a variety of physical phenomena involving light-matter interaction\cite{QuantumOpticsBooks}. One of these phenomena is the superradiance phase transition, which occurs when the interaction strength between the cavity and an ensemble of atom-like emitters exceeds a certain critical value and the different subsystems form a strongly correlated thermal equilibrium state \cite{Hepp,Wang}. In addition to its importance in representing a new regime of light-matter interaction, the strongly correlated state would be of the Greenberger-Horne-Zeilinger type, and it could serve as a resource for quantum technologies such as precise measurement or clocks.

The superradiance phase transition has been analyzed and debated theoretically for over 40 years \cite{Carmichael,Rzazewski,Emary,Nataf,Ashhab2010,Viehmann,Bakemeier,Xu,Vukics,Ashhab2013}, and recent experiments have started to observe evidence of the phase transition \cite{Baumann,Scalari,Baden}.

The recent emergence of superconducting qubits and resonators has enabled circuit-QED systems to access parameter regimes that were inaccessible with other systems \cite{Devoret,Niemczyk,FornDiaz2010,FornDiaz2016,YoshiharaDSC,YoshiharaSpectra}. In particular, the coupling strength of a single qubit to a cavity can now be made comparable to the bare energies of qubit and cavity excitations. Unconventional spectra characteristic of the corresponding highly correlated states have been observed in the experiments reported in Refs.~\cite{YoshiharaDSC,YoshiharaSpectra}. Another recent experiment coupled thousands of qubits to a single cavity, achieving an effective coupling strength that is not far from the theoretically predicted critical value \cite{Kakuyanagi}.

As superconducting qubits are in a sense macroscopic artificial atoms, they have more tunable parameters and larger parameter fluctuations than natural atoms. This situation allows us to explore novel parameter settings but at the same time requires us to take into consideration possibly very large parameter fluctuations. As phase transitions are abrupt changes that take place when a combination of system parameters satisfies some criticality condition, one question that arises is whether the sharpness of the phase transition is preserved when parameter fluctuations are taken into consideration. Some related questions have been discussed recently in the literature \cite{Eastham,Larson2009,Strack}. A recent study also investigated possible ways to mitigate the effect of parameter fluctuations in the time-domain superradiance that occurs when the emitters are initially excited and subsequently allowed to emit photons into the cavity \cite{Lambert,Yukalov}.

In this work, we address the question of whether a superradiance phase transition exists and how its condition is affected by parameter fluctuations in the Dicke model, where a large number of qubits interact with a single cavity. The model that we study is based on typical superconducting circuit-QED systems, which include the so-called bias term that is usually absent in traditional cavity-QED systems involving for example natural atoms in an optical cavity.

The remainder of this paper is organized as follows. In Sec.~\ref{Sec:Model} we present the basic model and its ground state in the absence of fluctuations, hence introducing some basic ideas of the superradiance phase transition. In Sec.~\ref{Sec:DeltaAndG} we consider the effect of including fluctuations in the qubit gaps and the qubit-cavity coupling strengths. In Sec.~\ref{Sec:AllParameterFluctuations} we include fluctuations in the bias points. In Sec.~\ref{Sec:FiniteTemperature} we consider the effect of thermal fluctuations on the phase transition condition. After a brief discussion of typical experimental parameters and a possible measurement procedure in Sec.~\ref{Sec:ExperimentalConsiderations}, we conclude with some final remarks in Sec.~\ref{Sec:Conclusion}.

\section{Model system and its behavior in the absence of fluctuations}
\label{Sec:Model}

Let us consider a system with $N$ qubits coupled to a single harmonic oscillator, and let us (as a first step) assume that the qubit parameters are identical. The Hamiltonian is given by
\begin{equation}
\hat{H} = - \frac{\Delta}{2} \sum_{i=1}^N \hat{\sigma}_x^{(i)} + \hbar\omega \left( \hat{a}^\dagger \hat{a} + \frac{1}{2} \right) + g \sum_{i=1}^N \hat{\sigma}_z^{(i)} (\hat{a}+\hat{a}^\dagger),
\end{equation}
where $\Delta$ is the qubit gap, $\omega$ is the cavity's characteristic frequency, $g$ is the coupling strength between a single qubit and the cavity, the operators $\hat{\sigma}_{\alpha}^{(i)}$ (with $\alpha=x,y,z$) are the Pauli operators of qubit $i$, and $\hat{a}$ and $\hat{a}^{\dagger}$ are, respectively, the annihilation and creation operators of the cavity.

In the absence of parameter fluctuations, it is natural to define the collective spin operators
\begin{equation}
\hat{S}_{\alpha}=\sum_{i=1}^N \frac{\hat{\sigma}_{\alpha}^{(i)}}{2},
\end{equation}
which obey the standard spin commutation relations up to the factor $\hbar$, which we have not included in the definition of $\hat{S}_{\alpha}$, i.e.~$\left[\hat{S}_{\alpha},\hat{S}_{\beta}\right]=i\varepsilon_{\alpha\beta\gamma} \hat{S}_{\gamma}$, where $\varepsilon_{\alpha\beta\gamma}$ is the Levi-Civita tensor. If we also define the operators
\begin{equation}
\hat{x} = \frac{ \hat{a} + \hat{a}^\dagger }{2}
\end{equation}
and
\begin{equation}
\hat{p} = -i \frac{ \hat{a} - \hat{a}^\dagger }{2},
\end{equation}
the Hamiltonian can be expressed as
\begin{eqnarray}
\hat{H} & = & - \Delta \hat{S}_x + \frac{\hbar\omega}{4} \left( \hat{a} + \hat{a}^\dagger \right)^2 - \frac{\hbar\omega}{4} \left( \hat{a} - \hat{a}^\dagger \right)^2 \nonumber \\ & & + 2 g \hat{S}_z (\hat{a}+\hat{a}^\dagger) \nonumber \\
& = & - \Delta \hat{S}_x + \hbar\omega \hat{x}^2 + \hbar\omega \hat{p}^2 + 4 g \hat{S}_z \hat{x}.
\end{eqnarray}
If we now take the classical limit, i.e. treat the spin $S$ as a continuous variable (with $\sqrt{S_x^2+S_y^2+S_z^2}=N/2$) and similarly treat $x$ and $p$ as classical position and momentum variables, and we look for the ground state of the system by minimizing the Hamiltonian, we find that this state obeys the relations
\begin{eqnarray}
p & = & 0, \nonumber \\
S_x & = & \frac{N}{2} \cos \theta, \nonumber \\
S_z & = & - \frac{N}{2} \sin \theta, \nonumber \\
\theta & = & \tan^{-1} \frac{4gx}{\Delta}, \nonumber \\
x & = & - \frac{2gS_z}{\hbar\omega}.
\end{eqnarray}
Combining these equations we obtain
\begin{equation}
x = \frac{2g}{\hbar\omega} \times \frac{N}{2} \times \frac{4gx/\Delta}{\sqrt{1+(4gx/\Delta)^2}}.
\end{equation}
For $4g^2N/(\hbar\omega\Delta)<1$, this equation has a single solution: $x=0$. When $4g^2N/(\hbar\omega\Delta)>1$, the equation has three solutions: $x=0$, which now is a local maximum of the energy and therefore does not correspond to the ground state, and
\begin{equation}
\sqrt{1+(4gx/\Delta)^2} = \frac{4g^2N}{\hbar\omega\Delta},
\end{equation}
or in other words
\begin{equation}
x = \pm \frac{\Delta}{4g} \sqrt{\left(\frac{4g^2N}{\hbar\omega\Delta}\right)^2 - 1}.
\label{Eq:CavityFieldInSuperradiantState}
\end{equation}

The above derivation describes the quantum phase transition between a normal state and a superradiant state, with the well-known phase transition condition $4g^2N/(\hbar\omega\Delta)=1$. If we go well above the transition point, we can approximate the above expression for $x$ as
\begin{equation}
x = \pm \frac{gN}{\hbar\omega}.
\end{equation}
In the classical case, the $N$ small spins that form the large $S=N/2$ all point in the same direction: along the z axis if the parameters correspond to the normal state or making an angle $\theta$ with the z axis if the parameters correspond to the superradiant state. Well above the transition point, $\theta$ can be approximated as
\begin{equation}
\theta = \pm \left( \frac{\pi}{2} - \frac{\hbar\omega\Delta}{4g^2N} \right).
\end{equation}
For later convenience we define the angle $\theta'$:
\begin{equation}
\theta' = \frac{\pi}{2} - |\theta|.
\end{equation}
This angle is equal to $\pi/2$ in the normal phase and approaches zero when we go deep into the superradiant phase.

We can now take the above results and infer from them the quantum mechanical ground state in the superradiant phase. A key point here is to include not only one of the two classical solutions, but rather a coherent superposition of the two:
\begin{widetext}
\begin{eqnarray}
\ket{0} & = & \frac{1}{\sqrt{2}} \Bigg[ \left( \cos\frac{\theta'}{2} \ket{L} + \sin\frac{\theta'}{2} \ket{R} \right) \otimes \cdots \otimes \left( \cos\frac{\theta'}{2} \ket{L} + \sin\frac{\theta'}{2} \ket{R} \right) \otimes \ket{-x_0} \nonumber \\ & & + \left( - \sin\frac{\theta'}{2} \ket{L} + \cos\frac{\theta'}{2} \ket{R} \right) \otimes \cdots \otimes \left( - \sin\frac{\theta'}{2} \ket{L} + \cos\frac{\theta'}{2} \ket{R} \right) \otimes \ket{x_0} \Bigg],
\label{Eq:ManyQubitSuperradiantGroundState}
\end{eqnarray}
where $x_0$ is given by the positive solution in Eq.~(\ref{Eq:CavityFieldInSuperradiantState}).

Note that the first excited state of the system has the same form as Eq. (\ref{Eq:ManyQubitSuperradiantGroundState}) but with a minus sign between the two terms (i.e. the two branches of the superposition):
\begin{eqnarray}
\ket{1} & = & \frac{1}{\sqrt{2}} \Bigg[ \left( \cos\frac{\theta'}{2} \ket{L} + \sin\frac{\theta'}{2} \ket{R} \right) \otimes \cdots \otimes \left( \cos\frac{\theta'}{2} \ket{L} + \sin\frac{\theta'}{2} \ket{R} \right) \otimes \ket{-x_0} \nonumber \\ & & - \left( - \sin\frac{\theta'}{2} \ket{L} + \cos\frac{\theta'}{2} \ket{R} \right) \otimes \cdots \otimes \left( - \sin\frac{\theta'}{2} \ket{L} + \cos\frac{\theta'}{2} \ket{R} \right) \otimes \ket{x_0} \Bigg].
\end{eqnarray}
\end{widetext}
The energy separation between the lowest two levels decreases exponentially with increasing $N$, such that the lowest two states are degenerate in the thermodynamic limit. As a result, one could say that the two branches of the superposition are the two possible broken-symmetry ground states in the superradiant phase.

Before proceeding any further, we examine the physical nature of the state given in Eq. (\ref{Eq:ManyQubitSuperradiantGroundState}). If we focus on one branch in the superposition, then each subsystem (i.e.~the oscillator or each one of the qubits) is in the ground state that corresponds to its own Hamiltonian plus a mean-field contribution from the other subsystems with which it interacts. In particular, the bias term of each qubit has a mean-field contribution from the finite average value of the cavity field, while the cavity feels a mean-field force that is the sum of many small contributions from all the qubits. This mean-field approximation of the state fails to capture more complex subsystem correlations, especially as we approach the transition point. However, it turns out to be a good approximation for investigating other aspects of the phase transition, including the transition point \cite{Vidal,Larson2016}. There is no reason to believe that fluctuations in the system parameters will modify this picture. We shall therefore use a similar mean-field approximation in our calculations below.

\section{Effect of fluctuations in $\Delta$ and $g$}
\label{Sec:DeltaAndG}

We have assumed above that all the qubits have the same values of $\Delta$ and $g$. We now ask what effect one can expect if one includes fluctuations in these parameters. In particular, would one still have the phase transition, or would it be smeared out?

As mentioned at the end of Sec.~\ref{Sec:Model}, we shall use a mean-field approach to investigate the effect of parameter fluctuations on the phase transition. We therefore take the Hamiltonian
\begin{equation}
\hat{H} = - \sum_{i=1}^N \frac{\Delta_i}{2} \hat{\sigma}_x^{(i)} + \hbar\omega \left( \hat{a}^\dagger \hat{a} + \frac{1}{2} \right) + \sum_{i=1}^N g_i \hat{\sigma}_z^{(i)} (\hat{a}+\hat{a}^\dagger)
\end{equation}
and follow a similar procedure as the one we used in Sec.~\ref{Sec:Model} to find the ground state under the classical approximation. In particular, we write the oscillator Hamiltonian in terms of the $x$ and $p$ variables. It is not very useful to define the collective spin operator here. But we can instead proceed by considering the expectation values of the individual spins and use these as the classical values of the spins. The ground state is described by the expectation values:
\begin{eqnarray}
p & = & 0, \nonumber \\
\langle \sigma_x^{(i)} \rangle & = & \cos \theta_i, \nonumber \\
\langle \sigma_z^{(i)} \rangle & = & - \sin \theta_i, \nonumber \\
\theta_i & = & \tan^{-1} \frac{4g_ix}{\Delta_i}, \nonumber \\
x & = & - \sum_{i=1}^{N} \frac{g_i \langle\sigma_z^{(i)}\rangle}{\hbar\omega}.
\end{eqnarray}
Combining these equations we find that in the ground state
\begin{eqnarray}
x & = & \sum_{i=1}^{N} \frac{g_i}{\hbar\omega} \times \frac{4 g_i x/\Delta_i}{\sqrt{1+(4 g_i x/\Delta_i)^2}} \nonumber \\
& = & x \sum_{i=1}^{N} \frac{4g_i^2}{\hbar\omega\Delta_i} \times \frac{1}{\sqrt{1+(4 g_i x/\Delta_i)^2}}.
\end{eqnarray}

Since at the phase transition point $x=0$, the condition for the phase transition is given by
\begin{equation}
\sum_{i=1}^{N} \frac{4g_i^2}{\hbar\omega\Delta_i} = 1.
\label{Eq:PhaseTransitionConditionWithDeltaAndGFluctuations}
\end{equation}
In other words, when the left-hand side of Eq.~(\ref{Eq:PhaseTransitionConditionWithDeltaAndGFluctuations}) is smaller than one the system is in the normal phase, and when the left-hand side of Eq.~(\ref{Eq:PhaseTransitionConditionWithDeltaAndGFluctuations}) is larger than one the system is in the superradiant phase. In the superradiant phase, the ground state has the same form as Eq. (\ref{Eq:ManyQubitSuperradiantGroundState}), but now each qubit has its own value of $\theta'$ (determined by the values of $g$ and $\Delta$ for that particular qubit), and $x_0$ is the value of $x$ obtained by solving the equation
\begin{equation}
\sum_{i=1}^{N} \frac{4g_i^2}{\hbar\omega\Delta_i} \times \frac{1}{\sqrt{1+(4 g_i x/\Delta_i)^2}} = 1.
\end{equation}
As a result, the sharp phase transition is preserved in the case where $\Delta$ and $g$ are nonuniform.

It is interesting to note here that if we include small fluctuations in $g$ and $\Delta$ around the average values $\bar{g}$ and $\bar{\Delta}$, we find that because $\sum_{i=1}^N g_i^2>N\bar{g}^2$ and $\sum_{i=1}^N \Delta_i^{-1}>N\bar{\Delta}^{-1}$, we can expect that the fluctuations in these parameters will make it somewhat easier to reach the transition point in the sense that the transition condition is satisfied when $4\bar{g}^2N/(\hbar\omega\bar{\Delta})$ is equal to a value that is somewhat smaller than one.

\section{Effect of fluctuations in the bias parameter $\epsilon$}
\label{Sec:AllParameterFluctuations}

In addition to the gap $\Delta$, superconducting qubits are often described by the bias parameter $\epsilon$, which we have not included in our analysis so far. This parameter appears as the coefficient of a term proportional to the single-qubit operator $\hat{\sigma}_z^{(i)}$ in the Hamiltonian. For flux qubits for example $\epsilon$ is set by the externally applied flux through the main qubit loop and the persistent current of the qubit. When the applied flux is set to $(n+1/2)\Phi_0$, where $n$ is an integer and $\Phi_0$ is the flux quantum, $\epsilon$ vanishes and the qubit is said to be biased at its symmetry point. For purposes of observing the bistability that we consider here, it is desirable to bias all the qubits at their symmetry points and eliminate the bias term in the Hamiltonian. However, since it is impossible to fabricate identical qubits, e.g.~all having exactly the same area, we must also consider fluctuations in $\epsilon$. 

The Hamiltonian is now given by
\begin{widetext}
\begin{equation}
\hat{H} = \sum_{i=1}^N \left( - \frac{\Delta_i}{2} \hat{\sigma}_x^{(i)} + \frac{\epsilon_i}{2} \hat{\sigma}_z^{(i)} \right) + \hbar\omega \left( \hat{a}^\dagger \hat{a} + \frac{1}{2} \right) + \sum_{i=1}^N g_i \hat{\sigma}_z^{(i)} (\hat{a}+\hat{a}^\dagger)
\end{equation}
\end{widetext}
The classical ground state is now described by the expectation values:
\begin{eqnarray}
p & = & 0, \nonumber \\
\langle \sigma_x^{(i)} \rangle & = & \cos \theta_i, \nonumber \\
\langle \sigma_z^{(i)} \rangle & = & - \sin \theta_i, \nonumber \\
\theta_i & = & \tan^{-1} \frac{\epsilon_i+4g_ix}{\Delta_i}, \nonumber \\
x & = & - \sum_{i=1}^{N} \frac{g_i \langle\sigma_z^{(i)}\rangle}{\hbar\omega}.
\end{eqnarray}
Combining these equations gives
\begin{equation}
x = \sum_{i=1}^{N} \frac{g_i}{\hbar\omega} \times \frac{(\epsilon_i+4 g_i x)/\Delta_i}{\sqrt{1+((\epsilon_i+4 g_i x)/\Delta_i)^2}}.
\label{Eq:SemiclassicalStateEquationWithEpsilon}
\end{equation}

Now we have a finite value of $x$ even for small values of $g$, because the qubits (which are now generally biased away from their symmetry points) give rise to a finite force felt by the cavity, shifting its equilibrium point. Let us, however, consider the point where
\begin{eqnarray}
0 & = & \sum_{i=1}^{N} \frac{g_i}{\hbar\omega} \times \frac{\epsilon_i/\Delta_i}{\sqrt{1+(\epsilon_i/\Delta_i)^2}},
\label{Eq:ZeroSolutionCondition}
\end{eqnarray}
and there should be such a point for a general distribution of $\epsilon_i$, assuming here that we have a global knob that simultaneously shifts all $\epsilon_i$ up or down. At this point, the mean-field contributions from the different qubits cancel each other completely at $x=0$, making $x=0$ an allowed solution for the classical ground state. If we assume that (1) the number of qubits $N$ is very large, (2) $\epsilon_i$ have a distribution that is symmetric about the average value and (3) $\epsilon_i$ is uncorrelated with other variables, then the special bias point that gives the $x=0$ solution is given simply by:
\begin{equation}
\sum_{i=1}^{N} \epsilon_i = 0.
\end{equation}

\begin{figure}[h]
\includegraphics[width=8.0cm]{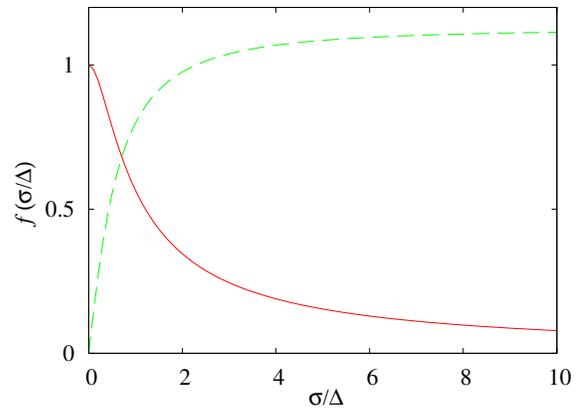}
\caption{Confluent hypergeometric function $f_1(\sigma/\Delta)=U\left(\frac{1}{2},0,\frac{\Delta^2}{2\sigma^2}\right)$ (dashed green line) and the function $f_2(\sigma/\Delta)=\frac{\Delta}{\sqrt{2}\sigma} U\left(\frac{1}{2},0,\frac{\Delta^2}{2\sigma^2}\right)$ (solid red line) as functions of the ratio $\sigma/\Delta$.}
\label{Fig:ConfluentHypergeometricFunction}
\end{figure}

We can then ask the following question: what is the strong-coupling condition when it is not possible to recover $x=0$ as the unique solution of Eq.~(\ref{Eq:SemiclassicalStateEquationWithEpsilon})? We observe that the left-hand side of Eq.~(\ref{Eq:SemiclassicalStateEquationWithEpsilon}) increases indefinitely with $x$ while the right-hand side saturates at a finite value when $x\rightarrow\infty$. The condition for having three solutions can then be re-expressed as the condition that the derivative of the right-hand side of Eq.~(\ref{Eq:SemiclassicalStateEquationWithEpsilon}) at $x=0$ is larger than 1 (which is the derivative of the left-hand side):
\begin{widetext}
\begin{equation}
\sum_{i=1}^{N} \left( \frac{4g_i^2}{\hbar\omega\Delta_i} \times \frac{1}{\sqrt{1+(\epsilon_i/\Delta_i)^2}} - \frac{4g_i^2 \epsilon_i^2}{\hbar\omega\Delta_i^3} \times \frac{1}{\left(1+(\epsilon_i/\Delta_i)^2\right)^{3/2}} \right) > 1,
\end{equation}
\end{widetext}
or in other words
\begin{equation}
\sum_{i=1}^{N} \left( \frac{4g_i^2}{\hbar\omega\Delta_i} \times \frac{1}{\left[1+(\epsilon_i/\Delta_i)^2\right]^{3/2}} \right) > 1.
\label{Eq:PhaseTransitionConditionWithEpsilonFluctuations}
\end{equation}
Note that $\epsilon_i$ are constrained by Eq.~(\ref{Eq:ZeroSolutionCondition}) above. If we assume that there are no qubit-to-qubit fluctuations in $\epsilon_i$, and we therefore set $\epsilon_i=0$, we recover the phase transition condition derived in Sec.~\ref{Sec:DeltaAndG}. The second factor inside the sum on the left-hand side of Eq.~(\ref{Eq:PhaseTransitionConditionWithEpsilonFluctuations}) decreases from 1 at $\epsilon_i=0$ down to zero in the limit $|\epsilon_i/\Delta_i|\rightarrow\infty$. We can therefore think of it as a factor $\alpha$ between zero and 1 that modifies the phase transition condition. A small value of $\alpha$ (which corresponds to large fluctuations in $\epsilon_i$) leads to the conclusion that larger values of $g_i$ are required in order to realize the superradiant phase. The crucial parameter here is the scale of variations in $\epsilon_i$ compared to the scale of $\Delta_i$. If the variations in $\epsilon_i$ are smaller than or comparable to $\Delta_i$, the transition point is shifted up by a factor of order one. On the other hand, if the variations in $\epsilon_i$ are much larger than $\Delta_i$ the above inequality (i.e.~the condition for the phase transition) can become orders of magnitude more difficult to reach than in the case with uniform $\epsilon_i$. To illustrate this point more clearly, let us assume that there are no fluctuations in $g$ or $\Delta$. We then obtain the condition for the transition point:
\begin{equation}
\frac{4g^2}{\hbar\omega\Delta} \times \sum_{i=1}^{N} \frac{1}{\left[1+(\epsilon_i/\Delta)^2\right]^{3/2}} = 1.
\end{equation}
If we assume that $\epsilon_i$ follow a Gaussian distribution with standard deviation $\sigma$, the sum is equal to
\begin{equation}
\frac{\Delta}{\sqrt{2}\sigma} U\left(\frac{1}{2},0,\frac{\Delta^2}{2\sigma^2}\right),
\end{equation}
where $U$ is the confluent hypergeometric function. This function is plotted in Fig.~\ref{Fig:ConfluentHypergeometricFunction}: it decreases from 1 to zero with increasing $\sigma/\Delta$, although the approach to zero is not very fast. In particular, the function is approximately equal to $\sqrt{2/\pi}\times\Delta/\sigma$ for large values of $\sigma/\Delta$.

As a result of the analysis in this section, we conclude that the phase transition exists also in the case of fluctuations in $\epsilon$. However, the transition point can now occur at a bias point where the qubits are on average away from their symmetry points. Furthermore, the phase transition condition is modified such that the coupling strength $g$ needed to satisfy the phase transition condition is larger than that obtained in the absence of fluctuations.

\section{Finite temperature}
\label{Sec:FiniteTemperature}

So far we have considered the quantum phase transition relating to the ground state. In this section we turn to the thermal phase transition between the normal and superradiant phases at finite temperatures.

\begin{figure}[h]
\includegraphics[width=9.0cm]{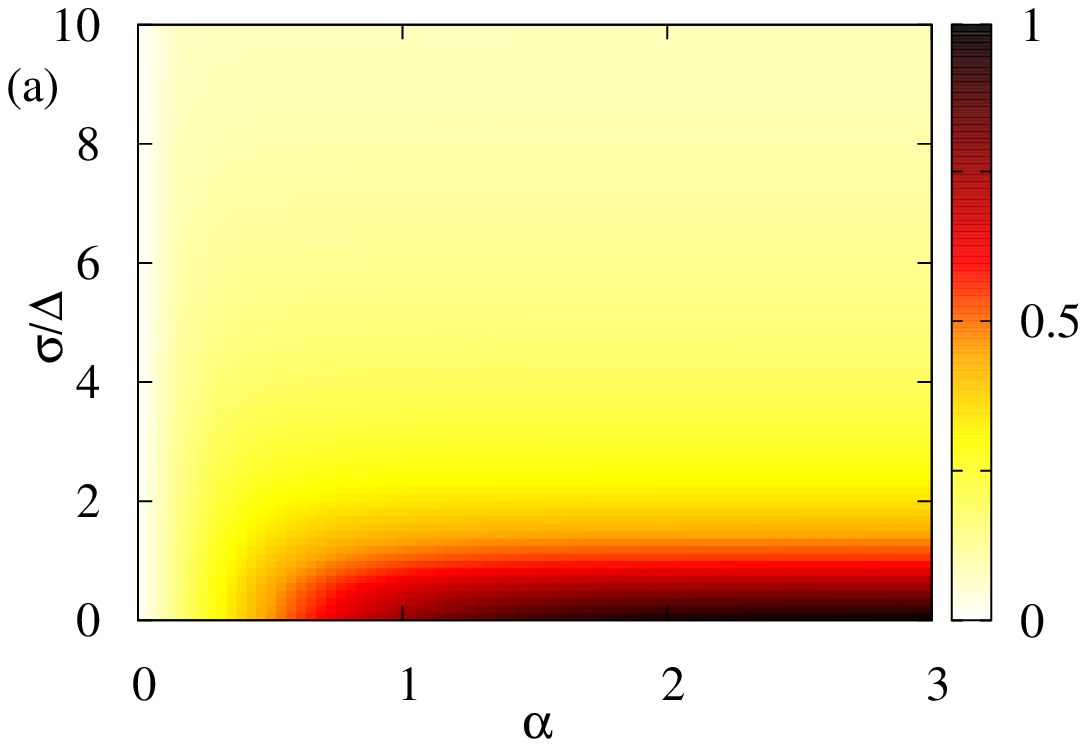}
\includegraphics[width=9.0cm]{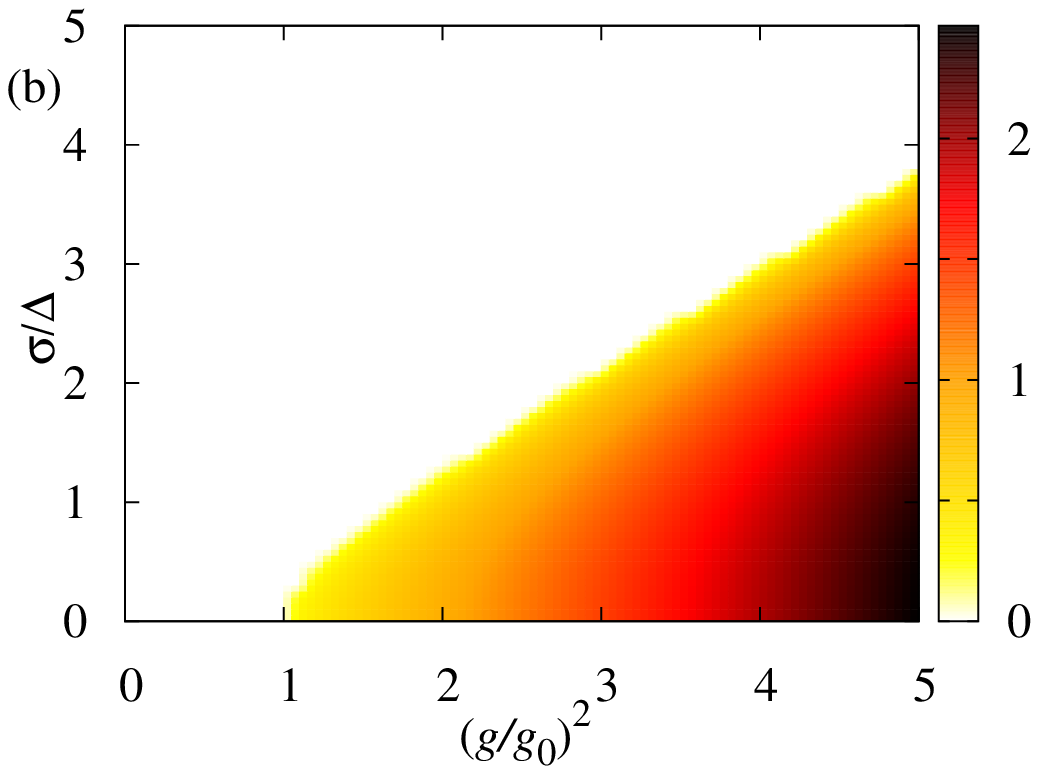}
\caption{(a) The sum in Eq.~(\ref{Eq:PhaseTransitionConditionWithEpsilonFluctuationsAndTemperatureSpecialCase}) as a function of $\alpha$ and $\sigma/\Delta$. This quantity approaches zero in the limits $\alpha\rightarrow 0$ and $\sigma\rightarrow\infty$, while it takes its maximum value 1 in the limit $\alpha\rightarrow\infty$ (i.e.~zero temperature) and $\sigma=0$ (i.e.~zero fluctuations). (b) The critical temperature $k_BT/\Delta$ as a function of coupling strength [expressed in the combination $(g/g_0)^2$ where $g_0=\sqrt{\hbar\omega\Delta/(4N)}$] and $\sigma/\Delta$. The critical temperature is zero at the boundary between the yellow and white regions, and inside the white region the system remains in the normal state even at zero temperature.}
\label{Fig:FiniteTemperature}
\end{figure}

Thermal fluctuations will modify the states of both the oscillator and the qubits. Increased temperature and hence increased fluctuations in a harmonic oscillator do not change the mean values of its $x$ and $p$ variables. In contrast, a qubit whose ground state is given by $\{\left\langle\sigma_x\right\rangle,\left\langle\sigma_y\right\rangle,\left\langle\sigma_z\right\rangle\}=\{\cos\theta_i,0,\sin\theta_i\}$ will have these values reduced at finite temperatures to $\{\left\langle\sigma_x\right\rangle,\left\langle\sigma_y\right\rangle,\left\langle\sigma_z\right\rangle\}=\{\cos\theta_i,0,\sin\theta_i\}\times\tanh[\hbar\Omega_i/(2k_BT)]$, where $\hbar\Omega_i=\sqrt{\Delta_i^2+\epsilon_i^2}$, $k_B$ is Boltzmann's constant and $T$ is the temperature. Equation (\ref{Eq:SemiclassicalStateEquationWithEpsilon}) then becomes
\begin{equation}
x = \sum_{i=1}^{N} \frac{g_i}{\hbar\omega} \times \frac{(\epsilon_i+4 g_i x)/\Delta_i}{\sqrt{1+((\epsilon_i+4 g_i x)/\Delta_i)^2}} \tanh\left[\frac{\hbar\Omega_i}{2k_BT}\right],
\label{Eq:SemiclassicalStateEquationWithEpsilonAndTemperature}
\end{equation}
with $\hbar\Omega_i=\sqrt{\Delta_i^2+(\epsilon_i+4g_i x)^2}$, and Eq.~(\ref{Eq:PhaseTransitionConditionWithEpsilonFluctuations}) becomes
\begin{widetext}
\begin{equation}
\sum_{i=1}^{N} \left( \frac{4g_i^2}{\hbar\omega\Delta_i} \times \frac{1}{\left[1+(\epsilon_i/\Delta_i)^2\right]^{3/2}} \tanh\left[\frac{\hbar\Omega_i}{2k_BT}\right] + \frac{2g_i^2}{\hbar\omega k_BT} \frac{(\epsilon_i/\Delta_i)^2}{1+(\epsilon_i/\Delta_i)^2} {\rm sech}^2\left[\frac{\hbar\Omega_i}{2k_BT}\right] \right) > 1.
\label{Eq:PhaseTransitionConditionWithEpsilonFluctuationsAndTemperature}
\end{equation}
This expression is not amenable to further analytic manipulation. However, one can understand how it affects the transition conditions by taking the special case where there are no fluctuations in $g$ or $\Delta$. In this case, Eq.~(\ref{Eq:PhaseTransitionConditionWithEpsilonFluctuationsAndTemperature}) can be expressed as
\begin{equation}
\frac{4g^2}{\hbar\omega\Delta} \times \sum_{i=1}^{N} \left( \frac{1}{\left[1+(\epsilon_i/\Delta)^2\right]^{3/2}} \tanh\left[\alpha\sqrt{1+\left(\frac{\epsilon_i}{\Delta}\right)^2}\right] + \alpha \frac{(\epsilon_i/\Delta)^2}{1+(\epsilon_i/\Delta)^2} {\rm sech}^2\left[\alpha\sqrt{1+\left(\frac{\epsilon_i}{\Delta}\right)^2}\right] \right) > 1,
\label{Eq:PhaseTransitionConditionWithEpsilonFluctuationsAndTemperatureSpecialCase}
\end{equation}
\end{widetext}
where $\alpha=\Delta/(2k_BT)$. In Fig.~\ref{Fig:FiniteTemperature}(a) we plot the sum on the left-hand side as a function of $\alpha$ and $\sigma$ (which is the standard deviation in $\epsilon$). As would be expected, this sum approaches zero in the limits $\alpha\rightarrow 0$ (i.e.~the high-temperature limit) and $\sigma\rightarrow\infty$ (i.e.~the large-fluctuation limit), because high temperatures and large fluctuations both favor the normal phase over the superradiant phase. In contrast, this factor takes its maximum value 1 when $\alpha\rightarrow\infty$ and $\sigma=0$, which is the special case of zero temperature and zero parameter fluctuations. In Fig.~\ref{Fig:FiniteTemperature}(b) we plot the critical temperature as a function of $4g^2N/(\hbar\omega\Delta)$ and $\sigma$. The superradiant phase can occur only when $4g^2N/(\hbar\omega\Delta)>1$, and even in this case a large value of $\sigma$ suppresses the critical temperature, such that the superradiant phase is easily destroyed by thermal fluctuations.

We finally note that when we ignore all parameter fluctuations and assume that all the qubit parameters are identical Eq.~(\ref{Eq:PhaseTransitionConditionWithEpsilonFluctuationsAndTemperature}) reduces to the well-known expression \cite{Hepp,Wang,Carmichael}
\begin{equation}
\frac{4g^2 N}{\hbar\omega\Delta} \tanh\left[\frac{\Delta}{2k_BT}\right] > 1.
\end{equation}

\section{Experimental considerations}
\label{Sec:ExperimentalConsiderations}

Although the superradiant state has been realized using a single flux qubit coupled to a superconducting LC circuit, reaching $g=7.63$ GHz and $4g^2/(\hbar\omega\Delta)\approx 10$ in Ref.~\cite{YoshiharaDSC}, such a strong coupling is still very difficult to realize in most experimental setups. Typical parameters for superconducting circuit-QED setups are: $\Delta\sim\hbar\omega\sim 2\pi\times$5 GHz, $g\sim 2\pi\times$10-100 MHz. These parameters give $4g^2/(\hbar\omega\Delta)\sim 2\times 10^{-5}$ - $2\times 10^{-3}$. One therefore needs $N\sim 10^3$ - $10^5$ in order to achieve the phase transition condition, even in the absence of parameter fluctuations. In particular, the recent experiment reported in Ref.~\cite{Kakuyanagi} had $g\sim 2\pi\times$15 MHz and $N=4300$, which gives $4g^2N/(\hbar\omega\Delta)\sim0.15$. These parameters are therefore somewhat below what is needed to realize the phase transition in the absence of fluctuations. As we have shown above, it is mainly fluctuations in $\epsilon$ that push the transition point towards higher values of $g$ and/or $N$. These fluctuations are typically on the order of 100 MHz, which is much smaller than $\Delta$ and should not lead to a dramatic increase in the required value of $g^2N$. As can be seen in Fig.~\ref{Fig:FiniteTemperature}, if we consider a value of $g/g_0$ between 1 and 2, the critical temperature is on the order of $\Delta/k_B$ and it decreases gradually to zero as $\sigma/\Delta$ increases and becomes comparable to 1. Achieving such temperatures in superconducting circuits is not very challenging, even though temperatures cannot go much lower than $\Delta/k_B$. We therefore expect that if for example an experiment realizes the parameters of Ref.~\cite{Kakuyanagi} with $g$ increased by a factor of 3-5, it could be possible to observe the superradiant phase.

The transition between the normal and superradiant states is usually studied by treating the coupling strength as a tunable parameter, with the superradiant state becoming the thermal equilibrium state for sufficiently strong coupling. If the coupling strength is tunable in a given experimental setup, one could vary it and measure the smallest frequency appearing in the absorption or transmission spectrum of the system. This frequency shrinks to zero at the transition point, which is a feature that can be used as a signature of the phase transition. It is common in superconducting circuits that the coupling strength is not tunable. Even in this case, where measuring the smallest excitation frequency in the spectrum would not allow one to identify the state of the system (i.e.~on which side of the transition point the system parameters lie), the full spectrum will generally allow a determination of the parameters \cite{YoshiharaSpectra}. Another possible technique for distinguishing between the two phases would be quantum state tomography. By suddenly shifting the qubit bias points away from the symmetry points, one obtains a situation where there is no exchange of excitations between the different subsystems, which enables measuring their individual states and hence performing quantum state tomography (see e.g.~Ref.~\cite{Ashhab2010}). For reliable results, this procedure would have to be performed within the coherence time that includes all decoherence processes, i.e.~the inverse of the sum of all the dephasing rates, which becomes experimentally challenging for a large number of qubits. Even if one cannot achieve the difficult task of many-qubit quantum state tomography, one could use a simpler technique that relies on the fact that in the superradiant phase the system will be equally likely to be in two different states. By performing the sudden shift of the qubit bias points away from the symmetry points and measuring only the state of the oscillator (many times), one can deduce whether the oscillator was always in the same state or in one of two different states. Hence this protocol can allow the identification of the bistable regime that corresponds to the superradiant phase. This procedure does not reveal any information about quantum coherence and therefore entanglement in the superradiant state, and even the qubit-oscillator correlations are deduced based on knowledge of the system parameters rather than experimentally measured correlations. However, since this technique relies on distinguishing between two coherent states in the oscillator, the only limiting factor for the measurement time is the oscillator's relaxation time.

Finally we note that the decoherence times $T_1$ and $T_2$ did not appear in our conditions for the phase transition, even though these parameters are closely related to noise and parameter fluctuations. The reason is that our study of the thermodynamic states deals with stationary states. As such, parameters related to the dynamics of the system will not necessarily appear in the results. In particular, $T_1$ and $T_2$ are generally determined by noise components at both zero frequency and the frequencies of the various transitions in the system, while thermodynamic properties are generally affected mainly by static fluctuations.

\section{Conclusion}
\label{Sec:Conclusion}

We have analyzed the effect of various combinations of parameter fluctuations on the superradiance quantum phase transition. Our results show that the phase transition is robust against these parameter fluctuations. Bias point fluctuations, which constitute one of the main limitations to coherence in superconducting qubit, have the most serious effect on the superradiance phase transition, because they can significantly increase the coupling strength required to realize the superradiant state. Nevertheless, a quantum phase transition is still expected to occur even in this case. Our results help guide future experiments to predict the conditions needed to observe the superradiance phase transition in realistic systems, and in particular superconducting circuit-QED architectures.

\section*{Acknowledgement}

We would like to thank M. Bamba, K. Kakuyanagi, Y. Matsuzaki, S. Saito and J. Vidal for useful discussions. This work was supported in part by Japan Society for the Promotion of Science (JSPS) Grant-in-Aid for Scientific Research (S) Grant No.~JP25220601.

\end{document}